\documentclass[a4paper,fleqn]{cas-dc}
\usepackage{subfigure}
\usepackage{caption}
\usepackage{lineno,hyperref}
\usepackage{url}
\usepackage[square,numbers]{natbib}
\usepackage{epsfig}
\usepackage[frozencache=true,cachedir=minted-cache]{minted}

\def\tsc#1{\csdef{#1}{\textsc{\lowercase{#1}}\xspace}}
\tsc{WGM}
\tsc{QE}

\begin{document}
\let\WriteBookmarks\relax
\def\floatpagepagefraction{1}
\def\textpagefraction{.001}

\shorttitle{Trends in Publishing Blockchain Surveys: A Bibliometric Perspective}    

\shortauthors{Hira Ahmad et al.}  

\title [mode = title]{Trends in Publishing Blockchain Surveys: A Bibliometric Perspective}  



%

\author[1]{Hira Ahmad}
\author[1]{Muhammad Ahtazaz Ahsan}
\author[1]{Adnan Noor Mian}\cormark[1]
\ead{adnan.noor@itu.edu.pk}
\ead[url]{www.itu.edu.pk}
\cortext[1]{Corresponding author}

\affiliation[1]{organization={Information Technology University},
            city={Lahore},
            country={Pakistan}}



\begin{abstract}
A large number of survey papers are being published in blockchain since the first survey appeared in 2017. A person entering into the field of blockchain is faced with the issue of which blockchain surveys to read and why? Who is publishing these surveys and what is the nature of these surveys? Which of the publishers are publishing more such surveys and what are the lengths of the published surveys? Which kind of survey is getting more citations? Which of the authors is collaborating on such surveys? etc. All these questions motivated us to analyze the trends in publishing blockchain surveys. In this paper, we have performed a bibliometric analysis on $801$ surveys or review papers published in the field of blockchain in the last approximately five years. We have analyzed the papers with respect to the publication type, publishers and venue, references, citations, paper length, different categories, year, countries, authors, and their collaborations and found interesting insights. To the best of our knowledge, this study is the first of its kind and hope to provide better understanding of the field.
\end{abstract}



\maketitle
\section{Introduction}
\label{sec:introduction}
Blockchain, first coined in $2008$ by Nakamoto \cite{nakamoto2008bitcoin}, has evolved as an emerging technology in the field of computer science. It is a decentralized, peer-to-peer network, in which nodes communicate with each other in a trust-less environment. 
The main structure of blockchain consists of blocks that are made up of transactions. The blocks are created through a process called mining in which a miner node collects, verifies, and adds new transactions into a block. The blocks are linked together in a chain-like structure through hashing. It has unique characteristics like making data tamper-proof and distributed ledger with no single point of failure. With such characteristics, it can be used in variety of applications, ranging from data security \cite{shrier2016blockchain}, data sharing \cite{fan2018medblock, zhang2018towards}, communications and networking \cite{liu2020blockchain}, and secure authentication \cite{kim2018secure} in various domains. Blockchain is also applied in other domains like the internet of things (IoT) \cite{dorri2017towards, dorri2016blockchain}, healthcare \cite{holbl2018systematic, mcghin2019blockchain}, fog and cloud computing \cite{tuli2019fogbus}, artificial intelligence (AI) \cite{alshamsi2021artificial, chamola2020comprehensive} or access control \cite{ourad2018using, ouaddah2017towards}, etc. 

A survey or review paper covers the state-of-the-art research techniques, challenges, opportunities, or future work in the specific area of knowledge. It also provides taxonomy or categorization of the literature and gives useful insights in the form of future work and challenges faced by the research community. To the best of our knowledge almost $9$ years after the blockchain was first introduced, the first survey paper was published in $2017$ and then in just about 5 years (till September $2021$) $801$ survey papers were published covering different domain areas. This is about $14$ survey papers in every month. This is about $5$ times higher than another field ``ad-hoc networks'' in which approximately $160$ survey papers were published in the same period of $2017$ to $2021$. With this high number of surveys, a person entering into the field of blockchain is faced with the issue of which blockchain surveys to read and why? Who is publishing these surveys and what is the nature of these surveys? Which of the publishers are publishing more such surveys and what are the lengths of the published surveys? Which kind of survey is getting more citations? Which of the authors is collaborating on such surveys? etc. With all these questions in mind, a natural inquiry and motivation into the trends of publishing in blockchain survey emerged.

In this paper, we have tried to answer above stated questions through a bibliometric analysis. A bibliometric analysis is a quantitative research evaluation approach that assesses the previous relevant scientific works based on quantitative indicators. It is performed to analyze and visualize the development of a scientific field. In this work, we performed bibliometric analysis on $801$ surveys or review papers published in the field of blockchain in the last approximately five years. To the best of our knowledge, this study is the first of its kind.

\textit{Organization of the Paper:}
The rest of the paper is organized as follows. We provide a brief overview of related work in Section \ref{sec:relatedwork} and then describe data collection and preprocessing methods in Section \ref{sec:dataandmeth}. We present bibliometric analysis in Section \ref{sec:bib_analysis} and finally we conclude our paper in Section \ref{sec:conclusion}.

\section{Related Work}
\begin{figure}[!h]
\centering
\includegraphics[scale=0.33]{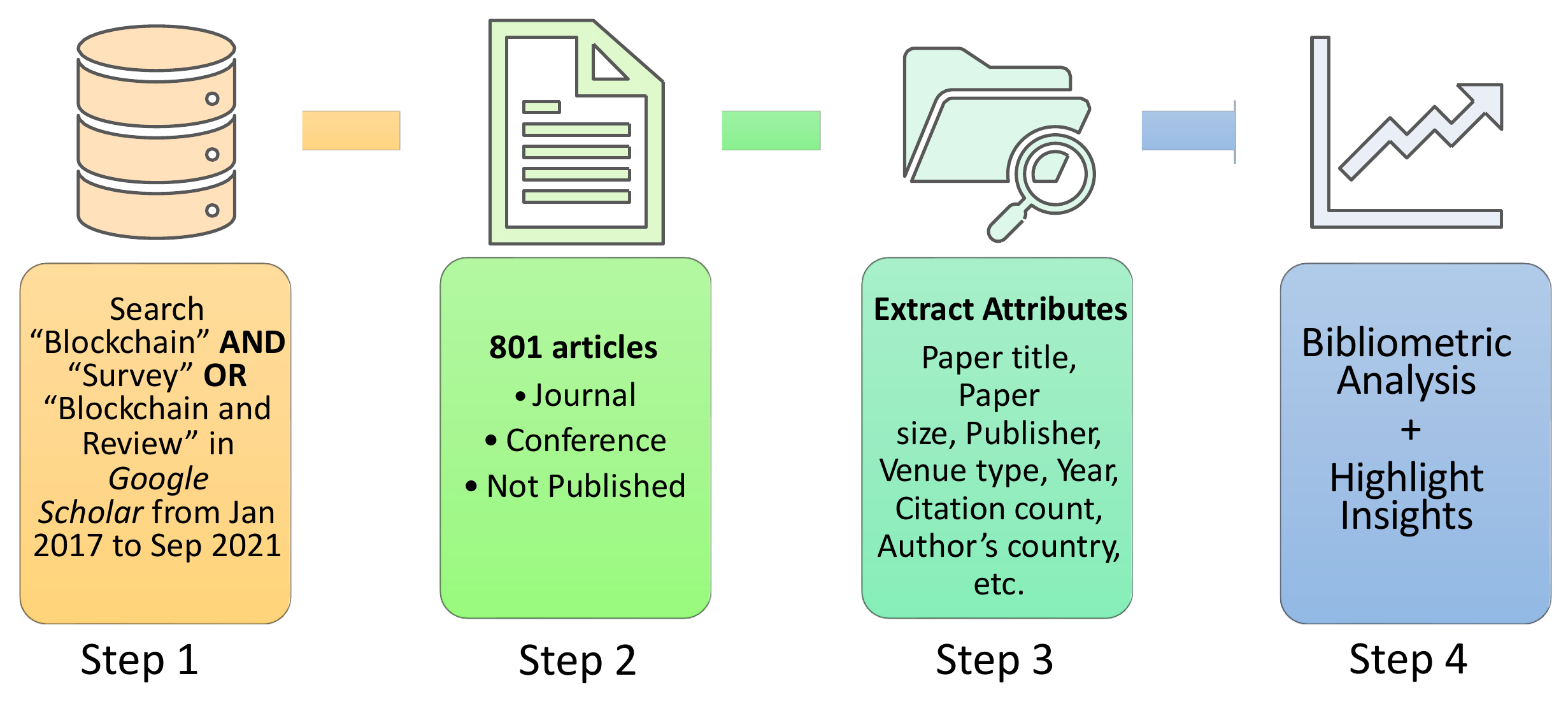}
\caption{Methodological steps}
\label{fig:fignew}
\end{figure}
\label{sec:relatedwork}

Firdaus et al. \cite{firdaus2019rise} used the term ``blockchain'' for collecting blockchain-related articles. They extracted $1119$ articles published during the period $2013$ to $2018$ from the \textit{Scopus} \cite{scopus} and found that (i) the future trend will be of solving IoT security issues, (ii) blockchain will be mostly used in healthcare, (iii) USA, China, and Germany have the most number of publications in blockchain, (iv) Singapore and Switzerland have fewer publications and many citations, (v) higher research collaborations means higher the publications except for Canada, India and, Brazil (vi) the keyword analysis showed that blockchain is used in various fields of research. Similarly, Guo et al.

\cite{guo2021bibliometric} has obtained $3826$ articles for blockchain published during $2013-2020$ from \textit{Web of Science} (WoS) \cite{WoS}, and used \textit{CiteSpace} \cite{chen2006citespace} and \textit{VOSviewer} \cite{van2010software} to extract publication trends, top-cited authors, highly cited journals, most-cited references, authors' network, top-productive countries and institutions, and emerging trends of blockchain. Zeng et al. \cite{zeng2018bibliometric} selected Ei Compendex \cite{Ei} and China knowledge infrastructure \cite{Cki} for blockchain-related literature to extract literature between January $2011$ and September $2017$. From both the sources, they analyzed the most productive authors and institutes, collaboration patterns among authors and institutes, and the emerging topics.

Ante \cite{ante2020smart} searched the term ``smart contract'' from WoS for analyzing $468$ articles having $20,188$ references with $15,714$ unique papers referenced. Smart contracts (SC) are simple programs that are stored on a blockchain and execute only when predetermined requirements are met. They automate the execution of a contract so that all members can instantly be sure of the outcome. This work has applied exploratory factor analysis for co-citation analysis to recognize six groups of research that are (i) blockchain networks development, (ii) blockchain and smart contracts for IoT, (iii) smart contract security, standardization, and verification, (iv) smart contracts and blockchain for the disruption of industries, (v) challenges of smart contracts, (vi) smart contracts and law. The work \cite{merediz2019bibliometric} claimed to be the first general bibliometric study of bitcoin literature that collected $1162$ papers during $2012$ to $2019$ from WoS and have found the leading authors, main research areas, countries with most publications, most productive authors, research clusters, and leading authors. Ante et al. in another publication \cite{ante2021blockchain} analyzed $166$ articles from WoS on the energy sector with blockchain, used exploratory factor analysis to find six research streams that are (i) energy market reform and change, (ii) blockchain for security and data sharing, (iii) energy management in scalable systems and smart grids, (iv) information sending across networks and its applications, (v) peer-to-peer energy micro-grids and (vi) blockchain technology potential. Social network analysis is applied to these streams to find the relationships and dependencies among them. The results showed that there was more than $71.6$\% of variance among the above mentioned streams.

M{\"u}{\ss}igmann et al. \cite{mussigmann2020blockchain} analyzed the articles from $2016$ to January $2020$ on the domain of logistics and supplychain management (LSCM) along with blockchain technology (BCT). The dataset was collected from $10$ databases including \textit{Scopus}, \textit{Google Scholar} (GS) and WoS, then applied refined data collection process which made the articles count to $613$. Authors then performed statistical analysis on the affiliations and collaborations between different authors, highlighted the keywords, and also performed a citation and a co-citation network analysis that helped to divide the existing work into five classes as (i) theoretical sense-making of BCT in LSCM, (ii) testing and conceptualizing blockchain applications, (iii) digital supplychain management, (iv) technical design of BCT applications for real-world LSCM applications, and (v) framing BCT in supplychains. Moosavi et al. \cite{moosavi2021blockchain} performed bibliometric analysis on the articles and book chapters collected from \textit{Scopus} for application of blockchain in supplychain to find out the important studies that let them define the supplychain areas and additional integrated technologies, main research groups, institutions, and countries. Tandon et al. \cite{tandon2021blockchain} selected $586$ articles from \textit{Scopus} covering the domain of management in blockchain and included $72$ countries, $273$ journals, $1016$ organizations, and $1284$ authors. Their findings are based on blockchain applications in particular managerial areas, e.g., finance and supplychain management. In their research work, they recognized four sub-categories of research as (i) policy and management, (ii) enablement of blockchain in management, (iii) multi-domain deployment, and (iv) incompetence of bitcoin.

For blockchain in the IoT domain, Kamran et al. \cite{kamran2020blockchain} conducted a bibliometric study on the dataset containing $151$ articles extracted from \cite{WoS}. The authors analyzed the yearly trends of publications, keyword analysis, the highest average citations per year, and the top listed venues. Anjum et al. \cite{anjum2020mapping} identified useful insights by performing a bibliometric study on blockchain and the healthcare domain. They identified yearly trends of publications, highest publications by authors, institutes, countries, and publishers from all over the world where the data was collected from \textit{Scopus} from January $2020$ to March $2020$.

All of the above related works mainly focus on the blockchain-based articles and do not discuss the specific trends in publishing blockchain surveys. In this work, we are only interested in looking into the trends in publishing surveys and reviews papers in blockchain.

\section{Data Collection and Pre-processing}
\label{sec:dataandmeth}

\begin{table*}
\centering
\caption{Different attributes, their types and brief description that are scrapped for performing bibliometric analysis.}
\label{tab:attrib}
\begin{tabular}{lll}
\hline
\multicolumn{1}{l}{\textbf{Attribute name}} & \multicolumn{1}{l}{\textbf{Type of Attribute}} & \multicolumn{1}{l}{\textbf{Description}}                       \\ \hline
Paper Title                                   & string                                          & Title of the published paper.                                   \\
Publishers                                    & string                                          & Publisher which accepted the paper for the publication.         \\
Type of Venue                                 & string                                          & Venue is a conference, journal, or an ArXiv.                    \\
Year                                          & integer                                         & Year in which the paper is published.                           \\
Citation count                                & integer                                         & Number of citations received by a paper.                        \\
References count                              & integer                                         & Number of reference papers that are cited in a paper.           \\
Author names                                  & string                                          & Names of first author and all other authors in a paper.       \\
Country                                       & string                                          & Country of the first author and all other authors in a paper. \\
Institute name                                & string                                          & Institute of the first author and all other authors.            \\
Paper length                                    & integer                                         & Length of the paper in number of pages. \\ \hline
\end{tabular}
\end{table*}

\begin{figure}[!h]
\centering
\includegraphics[scale=0.55]{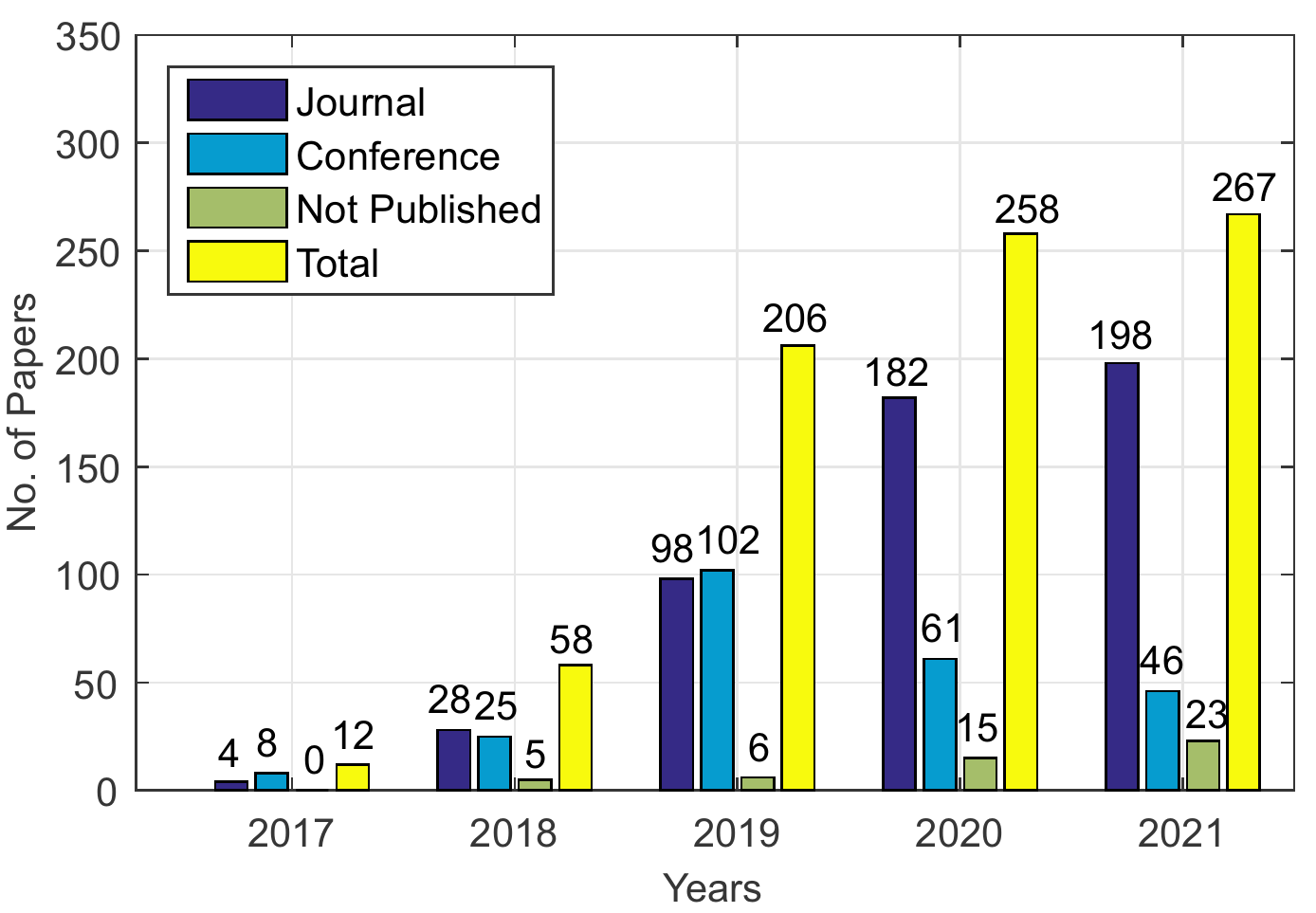}
\caption{Publication-wise papers}
\label{fig:fig2}
\end{figure}

Our methodology comprises of four major steps: (i) search the keywords from the database, (ii) collect the document, (iii) extract attributes from the preprocessed documents, and (iv) perform bibliometric analysis on the extracted attribute data. These steps are summarized in Figure \ref{fig:fignew}.

To collect our dataset, we scrapped the data from \textit{Google Scholar} (GS). For this purpose, we used \textit{Google Scholar's} advanced search option and queried the phrase which consists of the word ``$Blockchain \ AND \ Survey \ OR \ Blockch$- $ain \ AND \ Review$''. The purpose of choosing GS is because we can get a higher percentage of publications and citations over all the fields which are far greater than Scopus and WoS. In the field of computer science, GS has almost all the citations of Scopus and WoS, which makes it suitable over other databases \cite{martin2018google}. Moreover, GS is freely and easily accessible. We selected those papers which are published from January $2017$ to September $2021$. After applying our query, we downloaded papers ($801$) and collected different attributes of each paper, as shown in Table \ref{tab:attrib}. These attributes include paper title, publisher which has accepted the paper for publication, type of paper (conference, journal, or not published but available on arXiv), year in which the paper was published, count of citations a paper received till September $2021$, count of references (number of papers cited in particular survey or review paper), all authors, their country and institute names and, paper size (number of pages). After cleaning the data, we only consider $801$ survey papers.


\section{Bibliometric Analysis}
\label{sec:bib_analysis}

We now explore the attributes that we have mentioned above. For each attribute, we present the quantitative analysis and draw useful insights from them.

\subsection{Publication Type}
Publication type is used to classify the kind of paper that is published in a venue. The classes of publication type include journals, magazines, book chapters, conferences, letters, etc. We have considered only conferences, journals, and not published articles. Not published are those articles that have not gone through the formal process of peer review and publication and are available online on preprint servers, arXiv, TechRxiv and ResearchGate. Yearly statistics are shown in Figure \ref{fig:fig2}. It can be seen that there is an increase in the number of papers (survey and review) from $2017$ to $2021$. Usually a survey is written when a field of study has progressed sufficiently. Increase in number of surveys in blockchain means the field is quite dynamic and growing in a wider domain of applications. We also see from  Figure \ref{fig:fig2} that there are more number of journal papers ($510$) than the other two classes, i.e., conference ($242$) and not published ($49$). Note that the numbers in brackets are the cumulative sum of a particular publication type over the span mentioned above. Furthermore, the results in Figure \ref{fig:fig2} shows that number of journal and conference papers remain almost same till $2019$ and then the journal publications increased substantially as compared with that of conference publications. The recent trends in publishing blockchain surveys more in journals, is generally consistent with the publishing of surveys in other fields of computer science. 

\subsection{Publishers and venues}
\begin{figure}[!h]
\centering
\includegraphics[scale=0.5]{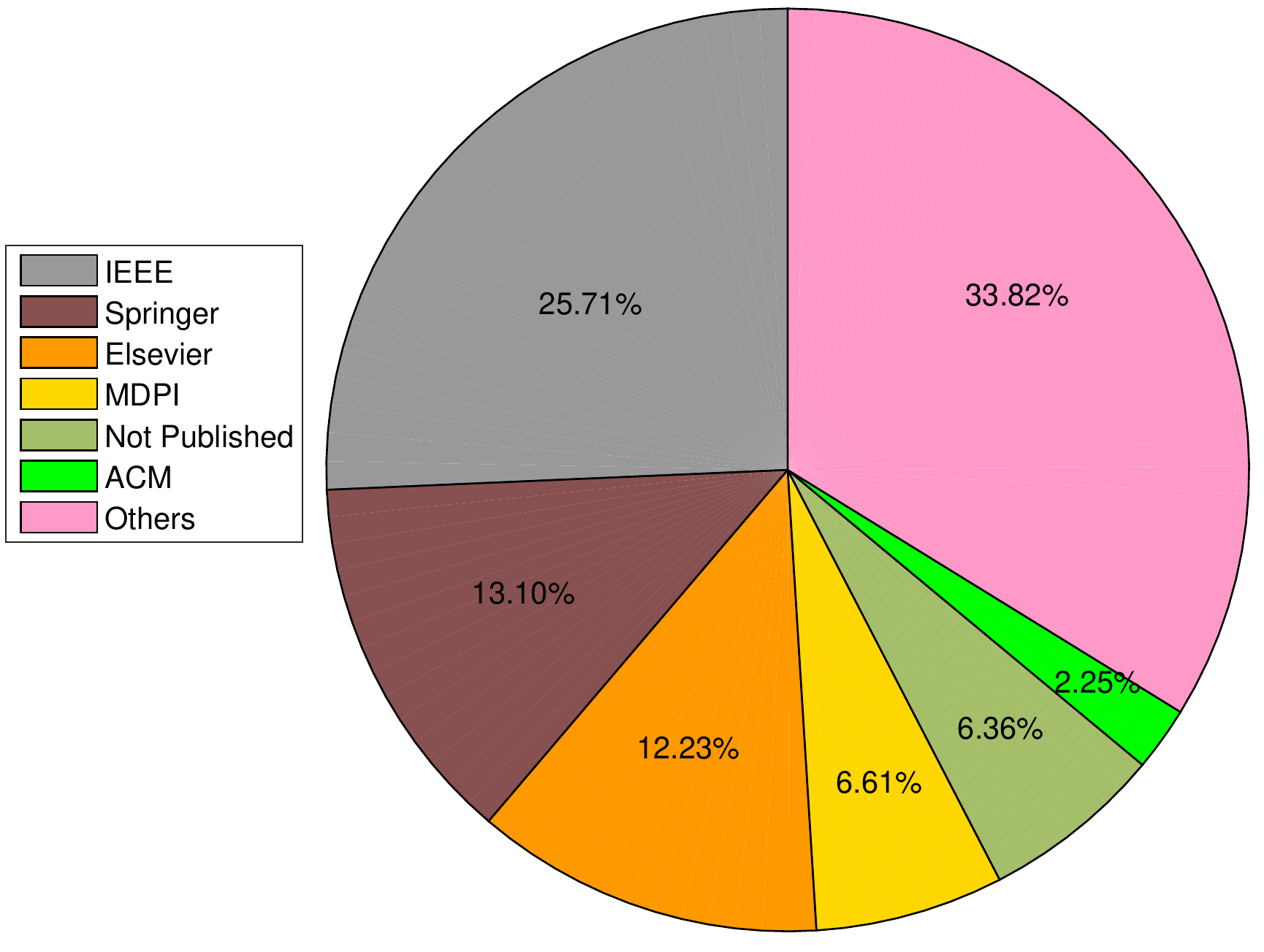}
\caption{Percentage of papers published under different venues}
\label{fig:fig3}
\end{figure}

A publisher is an organization that takes responsibility for a research paper's availability. In computer science, there are famous publishers like IEEE, Elsevier, ACM, etc. in which most of the researchers submit and publish their articles. We see that only $6.36$\% papers are not published and are available online in pre-print form. The distribution of overall published papers with different publishers is shown in Figure \ref{fig:fig3}. It is interesting to note that the maximum number of blockchain survey papers are published by IEEE ($25.71$\%) and the least by ACM ($2.25$\%). Springer, Elsevier, MDPI, not published (including ArXiv, TechRxiv, and ResearchGate) have $13.10$\%, $12.23$\%, $6.61$\%, and $6.36$\% respectively. In the ``others'' category, which form $33.82$\%, we combine all those publishers which have less than $5$ published papers.

\begin{table}[!h]
\centering
\caption{Number of published papers in top venues.}
\scalebox{0.7}{\begin{tabular}{llllcl}\hline
\textbf{\begin{tabular}[c]{@{}l@{}}Sr.\\ No.\end{tabular}} & \textbf{Venue} & \textbf{Publisher} & \textbf{Type} & \multicolumn{1}{l}{\textbf{Papers}} & \multicolumn{1}{c}{\textbf{Citations}} \\ \hline
$1$ & Access & IEEE & Journal & $57$ & $4287$ \\
$2$ & \multicolumn{1}{l}{\begin{tabular}[c]{@{}l@{}}Journal of Network and\\  Computer Applications (JNCA)\end{tabular}}  & Elsevier & Journal & $12$ & $791$ \\
$3$ & Applied Sciences & MDPI & Journal & $10$ & $231$ \\
$4$ & \multicolumn{1}{l}{\begin{tabular}[c]{@{}l@{}}Journal of Medical\\ Internet Research (JMIR) \end{tabular}}& \begin{tabular}[c]{@{}l@{}}JMIR\\ Publications\end{tabular} & Journal & $9$ & $143$ \\
$5$ & Sensors & MDPI & Journal & $8$ & $648$ \\
$6$ & \multicolumn{1}{l}{\begin{tabular}[c]{@{}l@{}}Emerging Telecommunications\\ Technologies (ETT) \end{tabular}} & Wiley & Journal & $8$ & $50$ \\
$7$ & CSUR & ACM & Journal & $7$ & $186$ \\
$8$ & ComST & IEEE & Journal & $7$ & $1422$ \\
$9$ & Computer Comm. & Elsevier & Journal & $6$ & $550$ \\
$10$ & IoT & IEEE & Journal & $6$ & $589$ \\
\hline
\end{tabular}}
\label{tab:toppublishers}
\end{table}

Furthermore, we have seen in the data that there are $206$ out of $801$ papers published by IEEE, among which $108$ are conference and $98$ papers are journal papers, Springer has published $58$ in conferences and $47$ in journals, ACM has also published $18$ survey papers, out of which $7$ are conference and $11$ are journals. Elsevier and MDPI both have published blockchain survey papers in journals only, with the count of $98$ and $53$, respectively. Statistics in terms of top number of publications in journals for different publishers and their venues is given in Table \ref{tab:toppublishers}. It is interesting to note that the journal IEEE Access has published the most number of surveys ($57$), whereas survey focused journals, ACM Computing Surveys (CSUR) and IEEE Communications Surveys \& Tutorial (ComST) has both published $7$ survey papers.


\subsection{Number of references}
\begin{figure}[!h]
\centering
\includegraphics[scale=0.54]{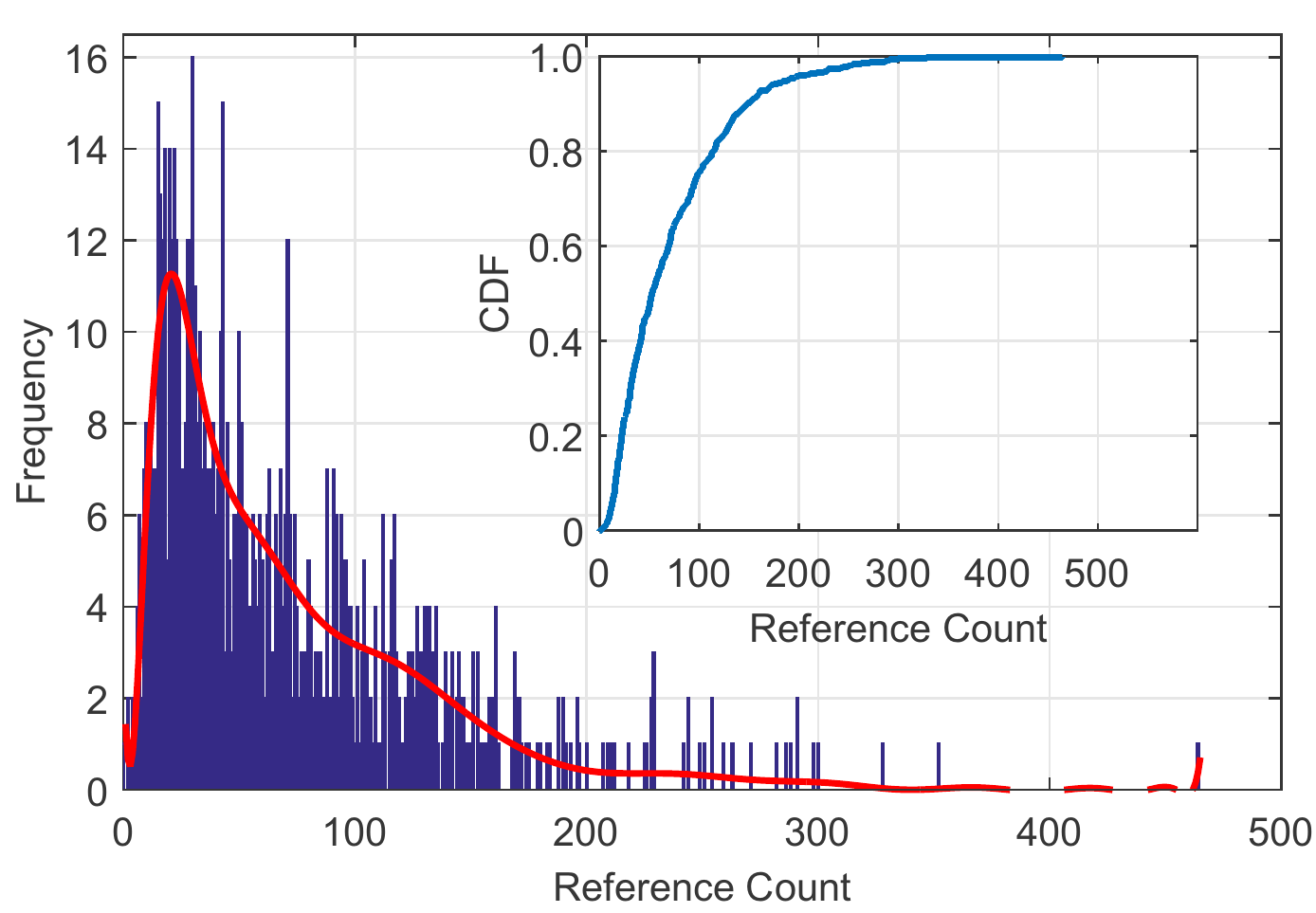}
\caption{Frequency distribution of number of references}
\label{fig:fig6}
\end{figure}

\begin{figure}[!h]
\centering
\includegraphics[scale=0.54]{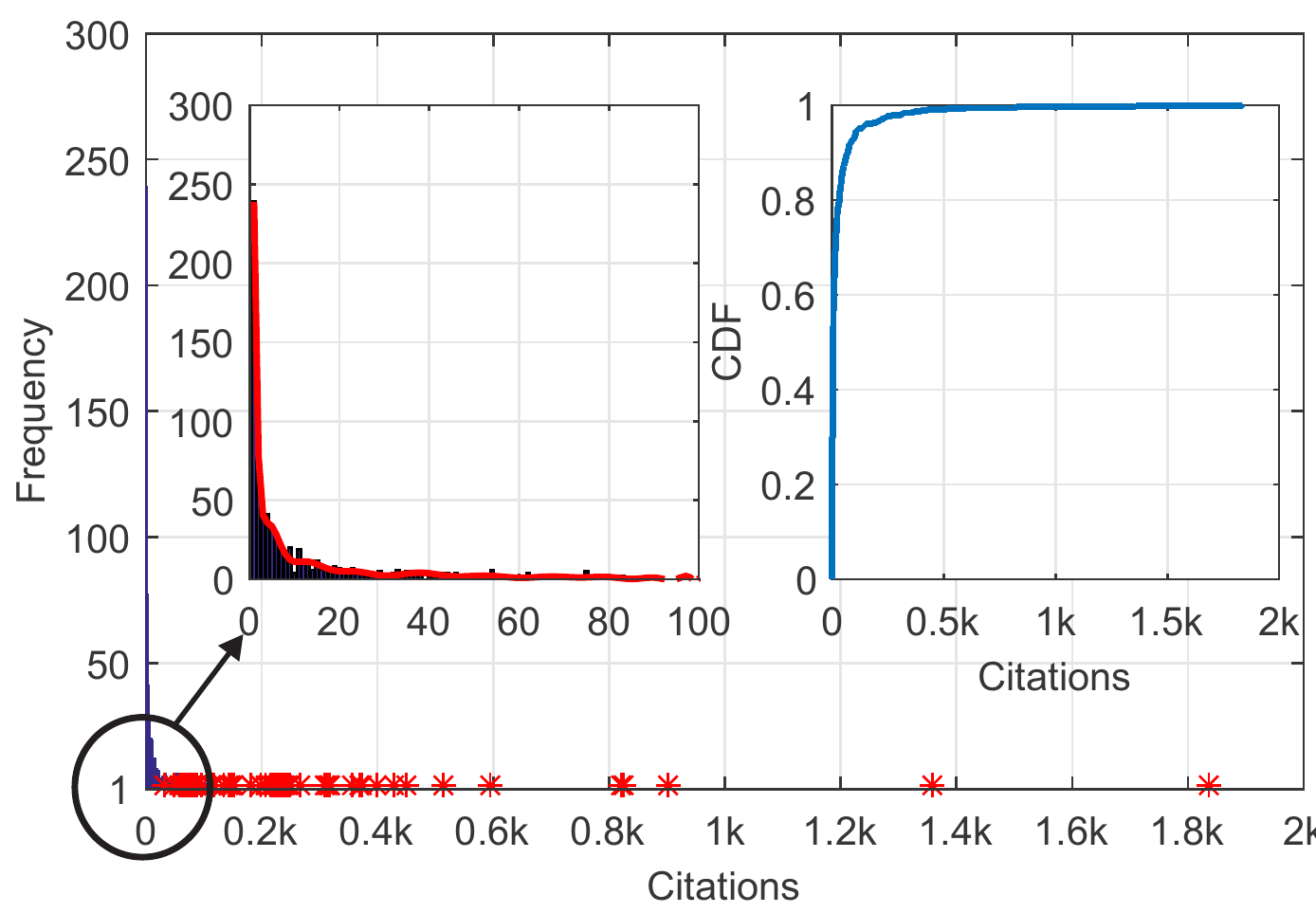}
\caption{Frequency distribution of number of citations}
\label{fig:fig5}
\end{figure}

\begin{figure}[!h]
\centering
\includegraphics[scale=0.54]{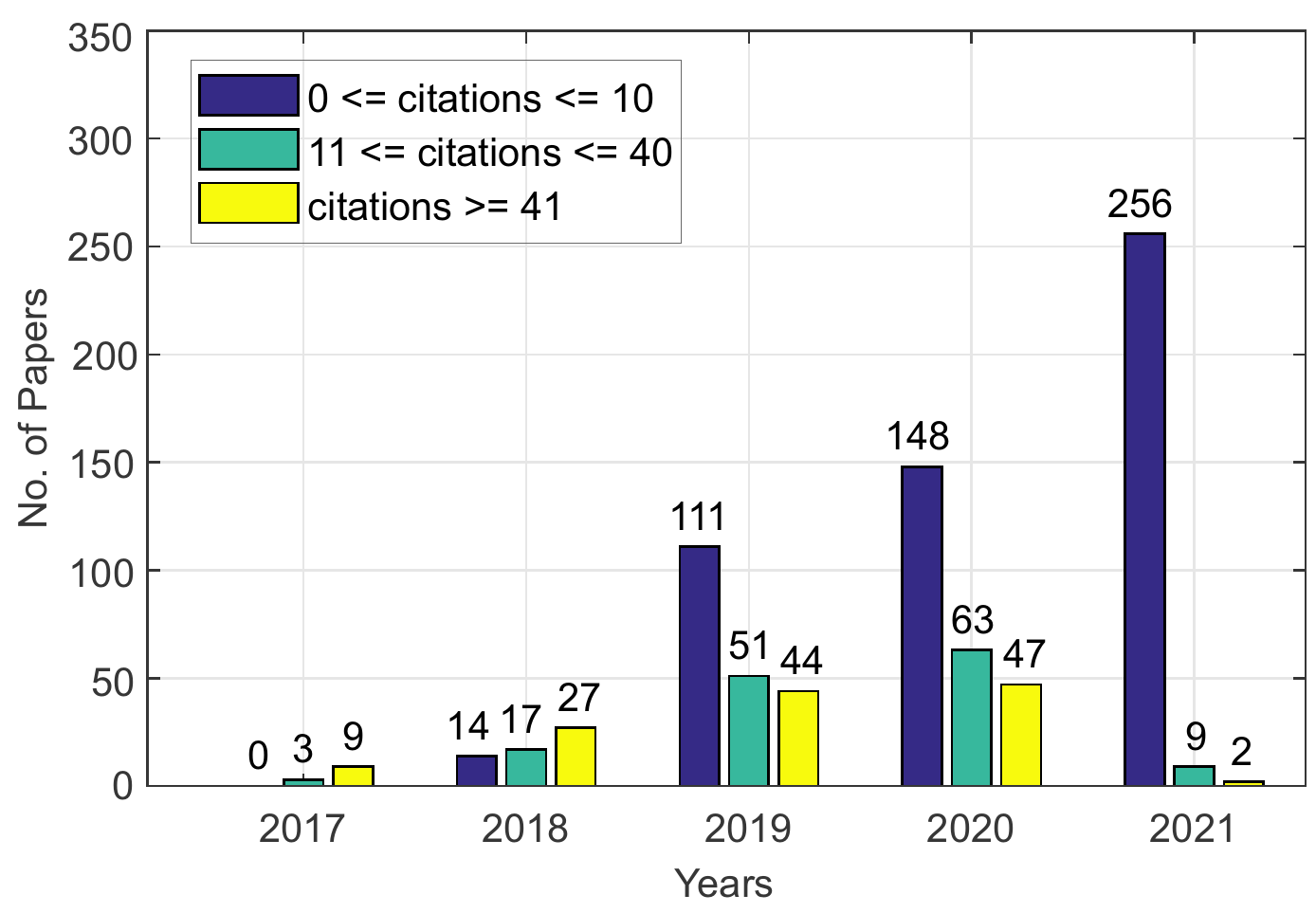}
\caption{Yearly trend in low, medium, and high number of citations }
\label{fig:citationsyear}
\end{figure}
References are those papers that are cited in a particular paper. Usually, the higher number of references in a survey paper shows an exhaustive literature review. Results for the number of references in blockchain surveys are shown in Figure \ref{fig:fig6}, where $x-{axis}$ represents the number of references used i.e., the count of papers cited in a survey and $y-{axis}$ shows the frequency i.e., number of papers which have used those references. From the plot in Figure \ref{fig:fig6} we see that generally there is a large number of papers with fewer references and fewer papers with a higher number of references. The subplot shows the cumulative distribution function(CDF) of the data.

\begin{figure}[!h]
\centering
\includegraphics[scale=0.54]{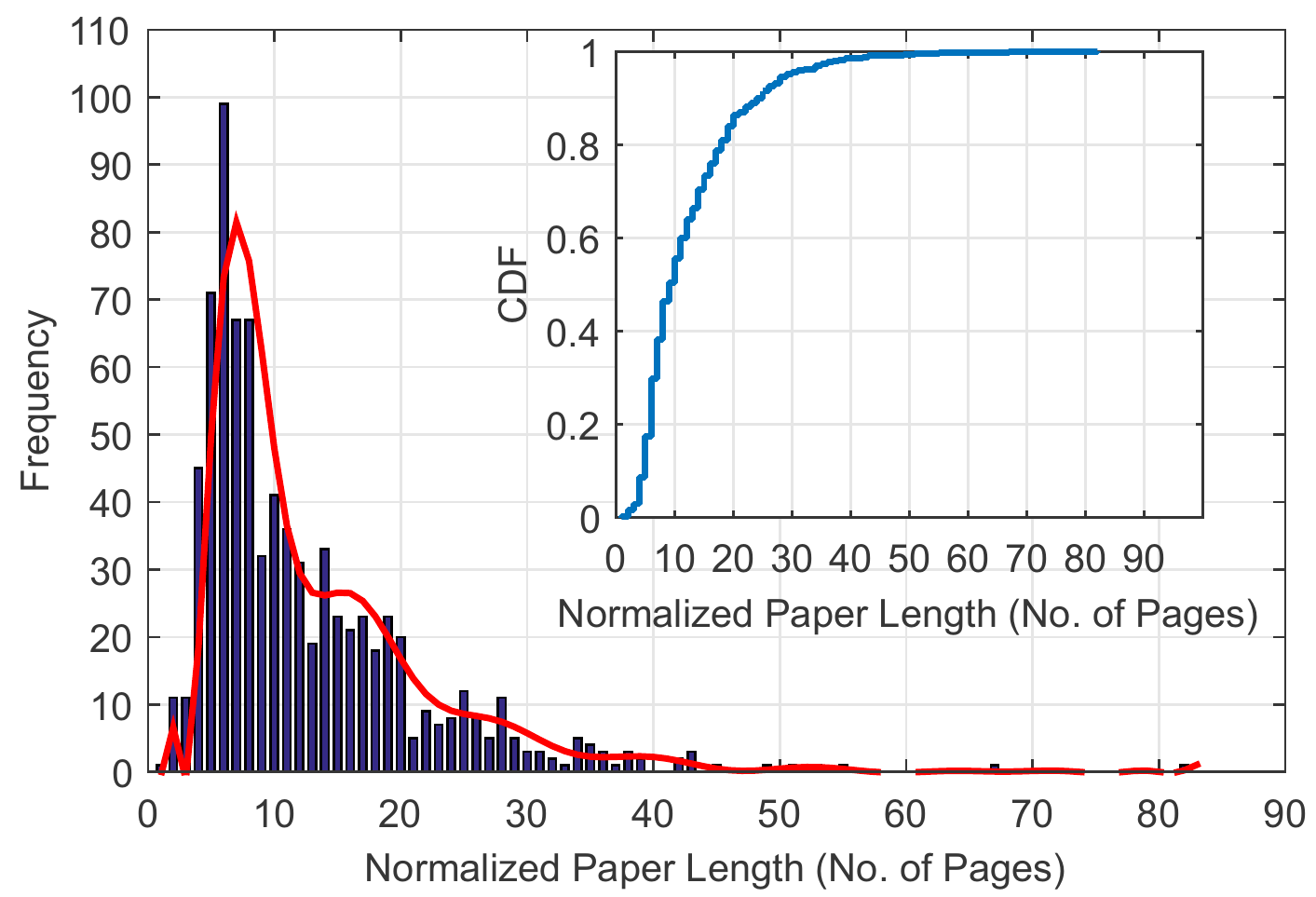}
\caption{Frequency distribution of length of survey papers}
\label{fig:fig7}
\end{figure}

Interestingly, half of the papers ($402$ out of $801$) have references up to $52$. There is only one paper \cite{casino2019systematic} that has a maximum of $464$ references. Moreover, the average reference count is approximately $71$. 



\subsection{Number of citations}
A paper gets cited when other authors refer to it in their papers. Results for citations are shown in Figure \ref{fig:fig5} where $x-{axis}$ shows the number of citations received by a paper and $y-{axis}$ represents frequency of that citation. There are $240$ out of $801$ papers that are not cited at all by any paper. We see a very less number of highly cited papers (shown by red in the plot). There is only one paper \cite{zheng2018blockchain} that has a maximum of $1835$ citations. The average number of citations is $31.57$. The area pointed by an arrow shows frequency of citations that lies within the range of $0$ to $100$. The CDF plot within Figure \ref{fig:fig5}, interestingly, shows that $83.89$\% of the papers have citations up to $40$. We have divided the total number of papers into three classes based on the number of citations they received. We assume paper less cited when they received $0$ to $10$ citations, medium cited when received $11$ to $40$ citations, and highly cited when received greater than $40$ citations. The results for these classes are shown in Figure \ref{fig:citationsyear}. Generally, the recent survey papers are expected to get more citations as they contain a more recent literature review. This fact can be seen in blockchain surveys in years $2017$, $2018$, and $2019$ but this does not seem to be evident in the latest years. Interestingly all of the blockchain papers published in $2017$ are got citations. 

\begin{figure}[!h]
    \centering
    \subfigure[Classification of surveys into different applications]{\includegraphics[scale=0.45]{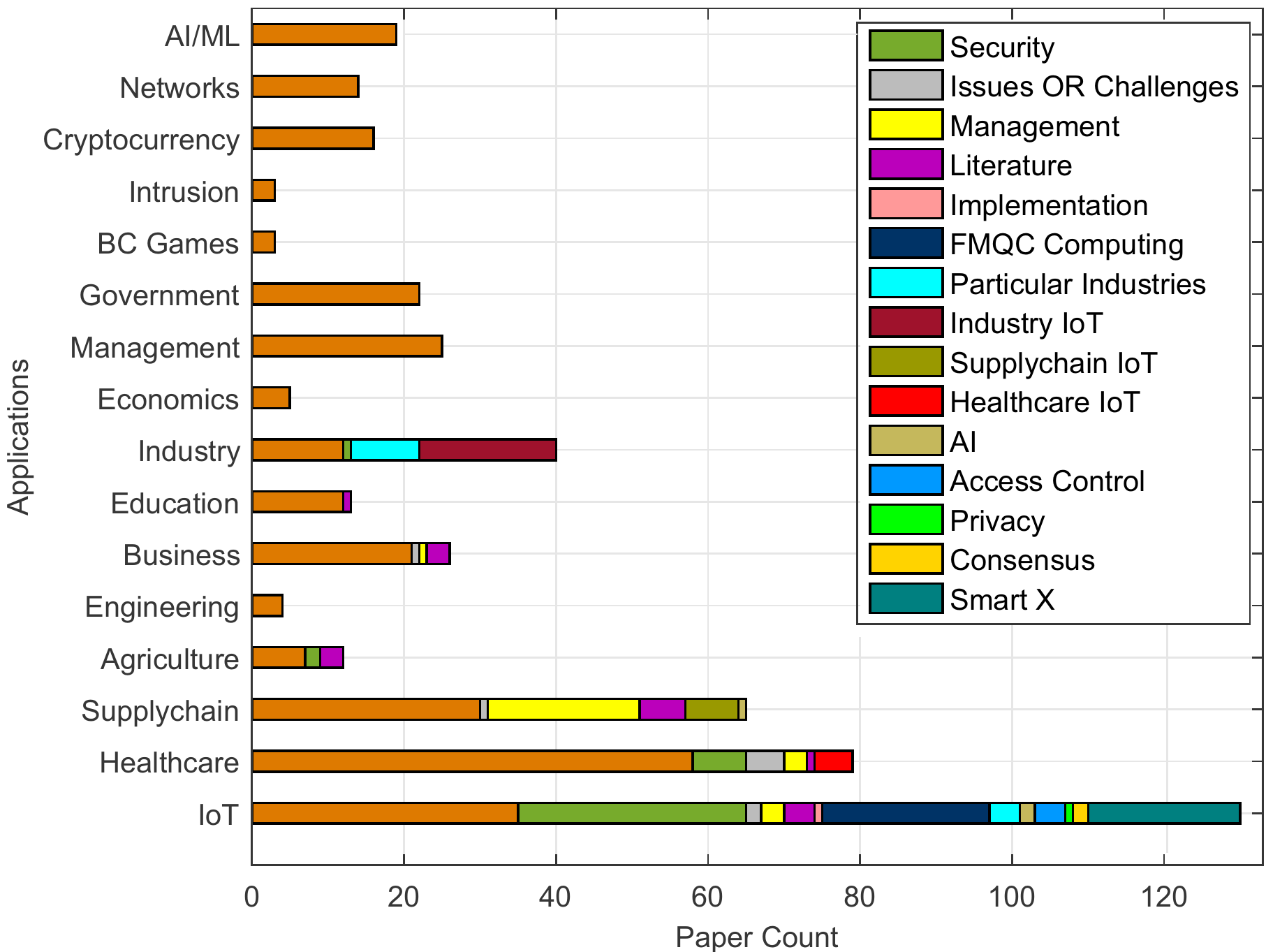}\label{fig:fig8}}
    \subfigure[Categorization survey published for blockchain features]{\includegraphics[scale=0.405]{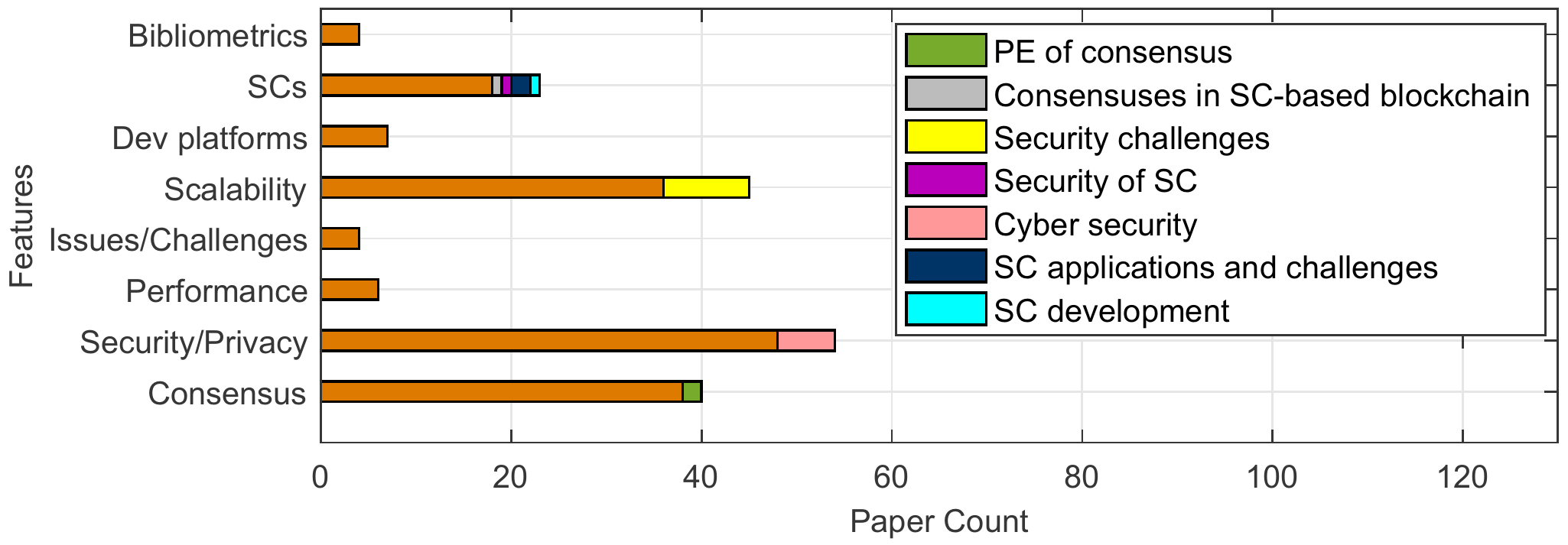}\label{fig:fig9}}
    \caption{Categorization of survey papers.}
    \label{fig:bc_categories}
\end{figure}

From Table \ref{tab:toppublishers}, we see that the number of citations for different venues. IEEE access has the highest number of citations ($4287$), whereas IEEE ComST has the second highest number of citations ($1422$). The citations per paper in $2$ years i.e., the total citations received by the journal divided by the total number of publications over the span of two years is defined as the impact factor \footnote{https://clarivate.com/webofsciencegroup/essays/impact-factor/} of a journal. From Table \ref{tab:toppublishers} we see that the number of citations of blockchain surveys per year of a journal is always quite larger than the impact factor of that journal. For example, in IEEE access this ratio is approximately $75$ which is much larger than its current impact factor of $3.367$ \footnote{https://ieeeaccess.ieee.org/}. This interesting fact can be due to the reason that the blockchain survey papers are getting a higher number of citations as compared to the other papers, including technical and survey both, published in the same venue in our period of study. We plan to study this fact more, including more factors, in our future work.


\subsection{Length of surveys}
Size of a paper is defined as its length in terms of number of pages. The larger size of a survey paper shows that authors have performed an extensive literature review. Statistical analysis for the size of each paper published is presented in Figure \ref{fig:fig7}. Generally, papers are published in two styles, i.e., in a single column or double column. For these results, we normalized the data by dividing the length of single column papers with the factor of $2$ and keeping the length of two column papers as it is. Generally, we consider papers as short survey papers that have size up to $6$ pages. We see in Figure \ref{fig:fig7} that a maximum of $99$ papers have a size of $6$ pages. Interestingly, we see in the data that there are $139$ survey papers having paper sizes less than $6$. The CDF plot in Figure \ref{fig:fig7} shows that $29.71$\% papers have a size up to $6$ pages. Any paper having $7\leq$ number of pages $\leq15$ is considered as medium-sized paper. We see in the data that $43.57$\% are medium-sized papers. Almost $86.39$\% papers have a size up to $20$ pages. Moreover, the average paper size is $12$ pages.




\subsection{Categorization of the surveys}
\label{subsec:classification}
Blockchain has been applied in many domains due to its vast applicability in healthcare, IoT, and smart cities, etc. Also, there are many papers to cover different features of blockchain, like smart contracts (SC), blockchain security, consensus or scalability, etc. Blockchain is also integrated with other domains like fog/mobile/quantum/cloud (FMQC) computing, or machine learning/artificial intelligence (AI), etc. We have classified the surveys into two main categories, (i) applications, (ii) features, based on their title strings. For this purpose, we removed stop words like is, of, to, on, for, etc. from the title strings and kept the keywords. For the sake of simplicity, we also removed the words like blockchain, survey, and review, from the title string just because they were present in each title string.  Further, we converted each word of a string into lowercase and also applied stemming (a process in natural language processing for obtaining the stem word, also called root word). The results  are shown in Figure \ref{fig:bc_categories}. 

The results for the first category (features) are shown in Figure \ref{fig:fig8}. It can be seen that IoT has the most number of papers ($130$). The field of IoT has also been applied in industries, healthcare, and supplychain domain. Most diverse sub-domains are also found in IoT as shown in the legend of Figure \ref{fig:fig8}. Also, we identified that in $2017$, there was only one  survey paper on using blockchain in IoT, and from $2018$ to $2021$, almost $160$ such survey papers were published, which comes out to be $3$ papers a month. The second highest number of survey papers are published in the healthcare domain ($79$). Although blockchain is mainly used for data integrity and security, interestingly we found that there are more survey papers with the title of IoT security ($30$) than healthcare security ($7$). The third most number of surveys are found for the Supplychain ($30$), and there are no survey papers that cover the security aspect of supplychain. Also, there are more survey papers on supplychain management ($20$) than IoT and healthcare management with the count of $3$ for both. There are very few papers on blockchain applications in government, economics, engineering, and education. The term Smart X (shown in legend of Figure \ref{fig:fig8}) covers smart homes, smart grids, and smart cities, and all of them count to $20$. Similarly, particular industries(shown in legend of Figure \ref{fig:fig8}) include pharmaceutical, shipping, construction, tourism, and vehicular industry.

The results for the second category (features) are shown in Figure \ref{fig:fig9}. In this category, we have further classified the keywords representing components of blockchain. Component can be consensus algorithm or scalability of blockchain, etc. Interestingly, it can be seen that the first, second, and third most number the papers are on the ``security/privacy'', ``scalability'', and ``consensus'' in blockchain, respectively. Consensus in blockchain is a method for maintaining a unified state of the ledger in a decentralized system. The most diverse surveys (covering more sub-domains of a particular domain) are conducted in the field of ``Smart contracts (SCs)''.  There are less number of surveys on ``bibliometrics'',  ``performance'' and ``development platforms'' which can be considered as future work for writing surveys.


\subsection{Countries, Authors, and their Collaborations}
Authors belonging to different countries collaborate with each other in writing survey papers. In this section first we provide the statistics on the contributions of different countries and then we present information about the authors and their collaborations.

\subsubsection{Statistics on country-wise publications}
\begin{figure}[!t]
\centering
\includegraphics[scale=0.31]{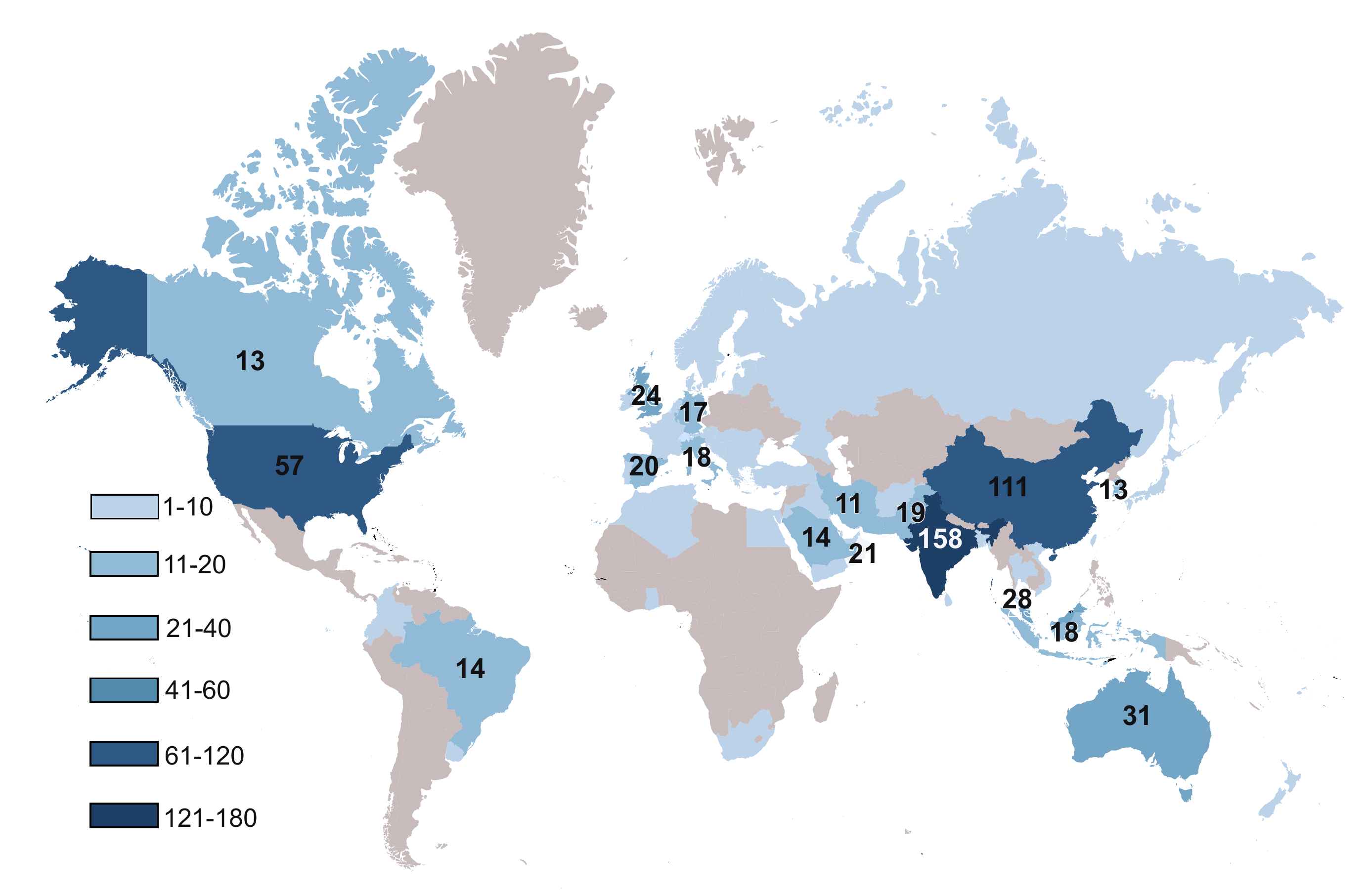}
\caption{Country-wise research papers published where most number of papers are published from China and India.}
\label{fig:fig11}
\end{figure}

The results for country-wise publications are shown in Figure \ref{fig:fig11}. Most number of papers are published by India and China, having count of $158$ and $111$, respectively. On the contrary, USA, Canada, and some of the European countries have published much lesser papers. We can say that most of the survey papers are published by Asian countries than Europe or USA. We do not show the number of publications of every country to make the diagram simple and only show the survey counts of those countries whose count is greater than $10$. Interestingly, it can be seen from Figure \ref{fig:fig11} that most of the countries have published less than $10$ surveys. It is also observed from the data that $20$ countries have published only one paper.

\begin{table}[!h]
\centering
\caption{Statistics on authors and their collaboration}
\begin{tabular}{lc}
\hline
\multicolumn{1}{c}{\textbf{Attribute}} & \textbf{Count} \\ \hline
Total unique authors in survey papers &$2556$ \\
Unique first authors &$776$ \\ 
\multicolumn{1}{c}{Unique other authors (other than first authors)} &$1853$ \\ 
Total collaborations between pair of authors &$4638$ \\ \hline
\end{tabular}
\label{tab:authstats}
\end{table}

\begin{figure*}[h]
    \centering
    \subfigure[Author's collaboration network, where $1$, $2$, $3$, $4$, and $5$ (in the legend) represents the number of papers in which authors have collaborated.]{\includegraphics[width=0.75\textwidth]{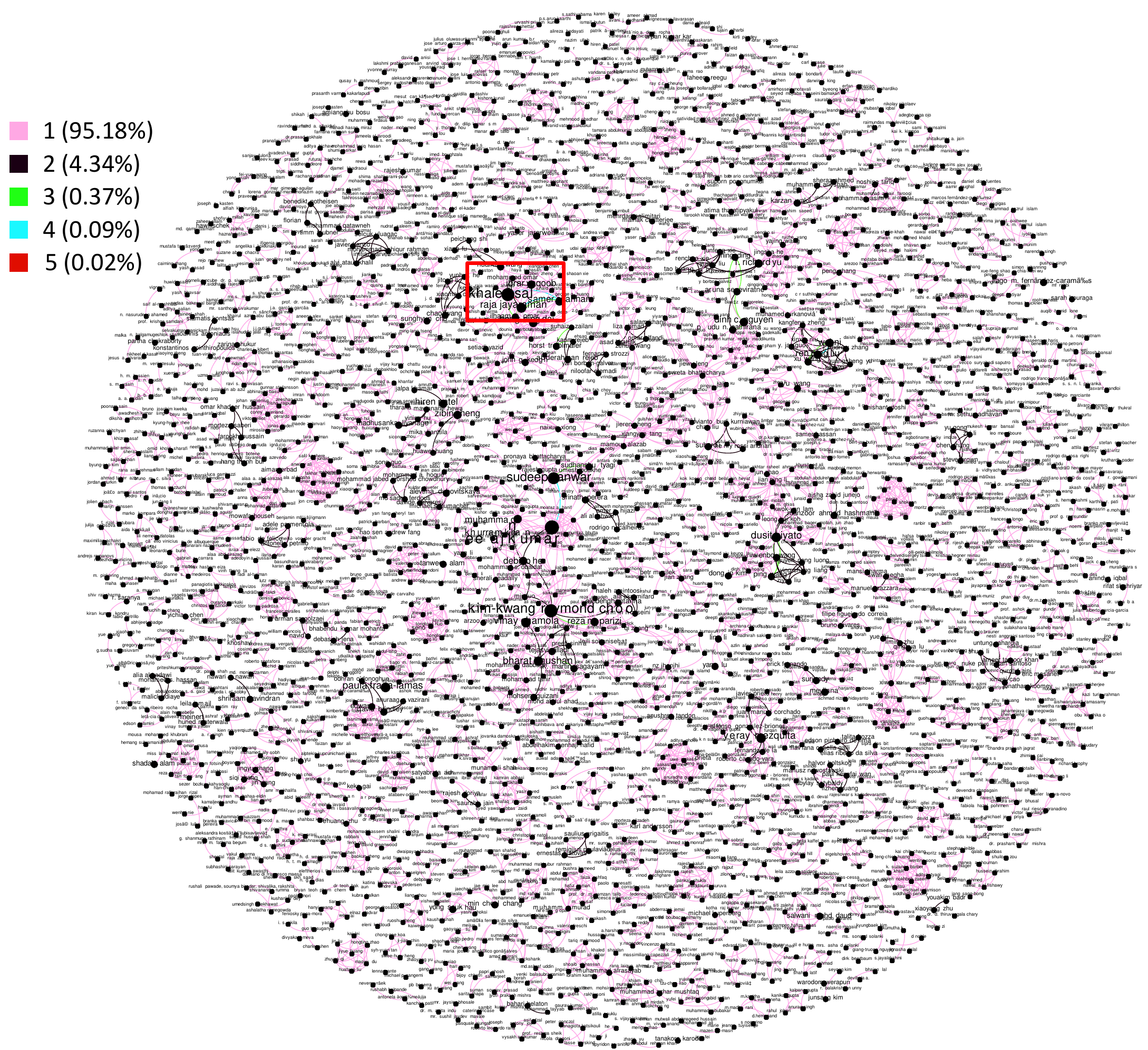}\label{fig:collaborations}}
    \subfigure[A portion where the multi-color links are visible.]{\includegraphics[width=0.5\textwidth]{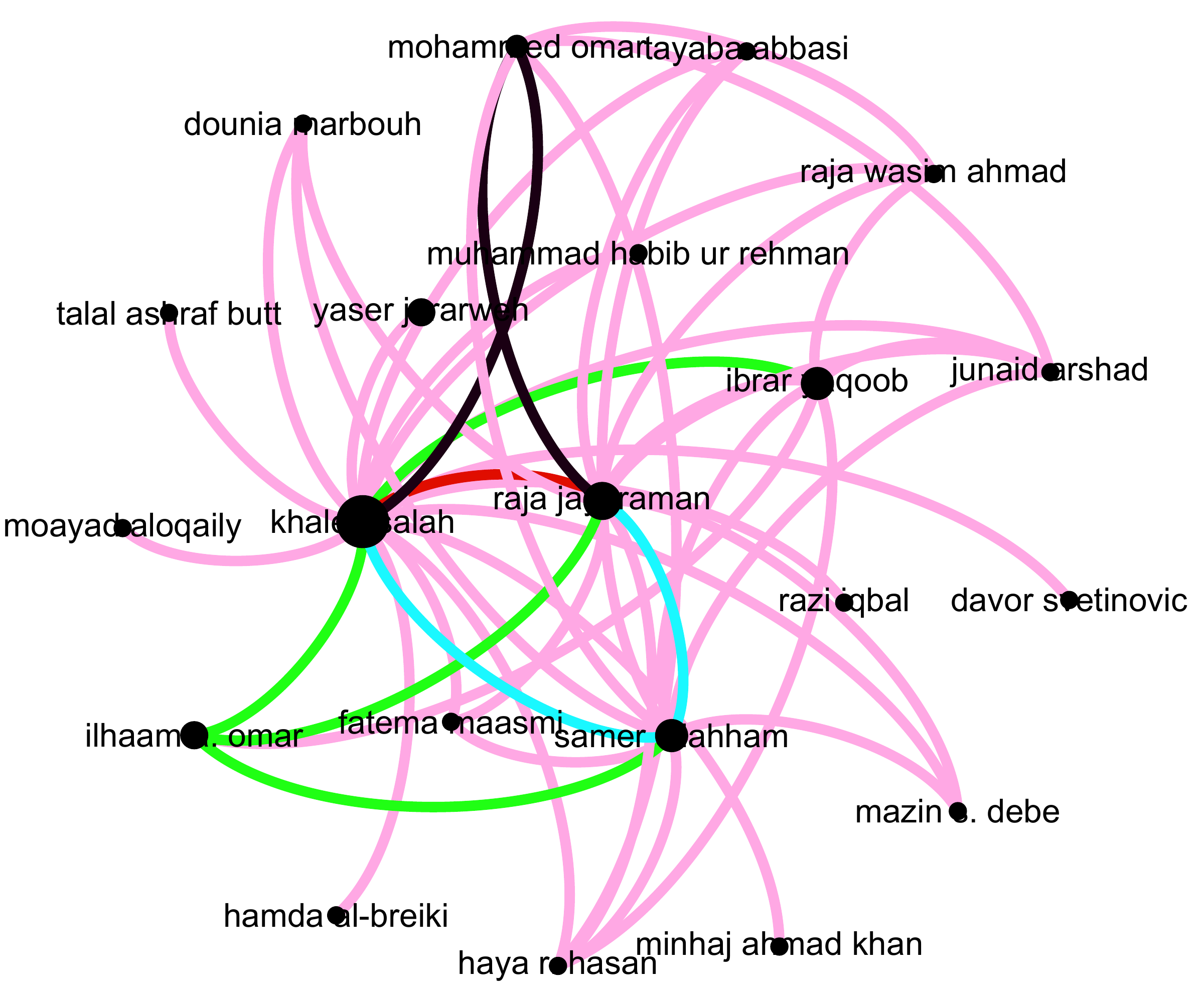}\label{fig:multi_links}}
    \caption{Authors and their collaborations diagram}
    \label{fig:collaborations_all}
\end{figure*}

\subsubsection{Authors' Collaborations}
We have identified the collaboration between different authors. To find the collaboration between authors, we have assumed that if any two authors are present in the same paper, they have collaborated with each other in that paper. To perform this, first of all, we identified all first authors. As a standard practice, all papers must have mentioned the name of the first author. Then, we collected the names of other authors as well. We then combined all first authors and second authors into a set and gave them a unique identifier. We then found the pair wise authors' collaboration using their identifiers. The overall statistics on the number of authors and the number of collaborations between them are provided in Table \ref{tab:authstats}. It can be seen in Table \ref{tab:authstats} that there are $2556$ unique authors. Statistics about total number of first authors, other authors, and number of pairs for collaboration can also be seen from the Table \ref{tab:authstats}.

\begin{table*}[!h]
\centering
\caption{Most cited papers and their domains}
\scalebox{0.9}{\begin{tabular}{cllll}
\hline
\multicolumn{1}{l}{\begin{tabular}[c]{@{}l@{}}
Sr.\\ No.\end{tabular}} & Papers & Year & Area of  Research & No. of citations \\ \hline
1 & Zheng et al. \cite{zheng2018blockchain} & 2018 & Challenges and opportunities & 1835 \\
2 & Khan and Salah  \cite{khan2018IoT} & 2018 & IoT security & 1358 \\
3 & Li et al.  \cite{li2020survey} & 2020 & Blockchain security & 927 \\
4 & Lin and Liao \cite{lin2017survey} & 2017 & Blockchain security & 838 \\
5 & Casino et al.  \cite{casino2019systematic} & 2019 & Applications & 831 \\
6 & Fernandez-Carames and Fraga-Lamas  \cite{fernandez2018review} & 2018 & IoT & 606 \\
7 & Hawlitschek et al. \cite{hawlitschek2018limits} & 2018 & Trust-free systems & 460 \\
8 & Mingxiao et al. \cite{mingxiao2017review} & 2017 & Consensus algorithms & 410 \\
9 & Panarello et al. \cite{panarello2018blockchain} & 2018 & IoT & 376 \\
10 & Sankar et al. \cite{sankar2017survey} & 2017 & Consensus algorithms & 320 \\ \hline
\end{tabular}}
\label{tab:topcitations}
\end{table*}

To create an author collaboration network, we used \textit{Gephi} \cite{bastian2009gephi}, which is open-source software for visualizing different kinds of networks.  \textit{Gephi} uses two separate files, i.e., node file and link file, for generating the network. The node file contains name of an author along with unique identifier is written. In link file, a link in the form of pair (source, target) of authors is written. These pairs must have those identifiers that are written in node file against the name of an author. The results for the author's collaboration are shown in Figure \ref{fig:collaborations}. Note that node size is proportional to the total number of papers in which the author participated as the first author or in any order like the second, third author, etc. It can be seen from Figure \ref{fig:collaborations} that most of the authors have collaborated with each other only in one paper, i.e., $95.18$\% of the total links are single paper collaboration links (as shown in light pink color). There are $4.34$\% links showing the author's collaboration in two papers (black color), $0.37$\% in three papers (green color), $0.09$\% links are showing collaboration in $4$ papers (cyan color) and $0.02$\% links are showing collaboration in $5$ papers (red color), respectively. The highlighted links between all proportions of collaborations are shown in Figure \ref{fig:multi_links}. It can be seen from Figure \ref{fig:multi_links} that the author named Khaled Salah has collaborated with different authors multiple times. For instance, author Khaled Salah has collaborated with: (i) Raja Jayaraman in five survey papers, (ii) Samar Ellahham in four survey papers, (iii) Ibrar Yaqoob and Ilhaam a. Omar in three survey papers, (iv) Muhammad Umar in two survey papers, (v) Muhammad Habib ur Rehman, Muhammad Al-Breiki and Davor Sretinovic, etc. one time. Author Neeraj Kumar has published a maximum of $7$ research papers. Overall, the author Khaled Salah contributed in $8$ different survey papers. Similar patterns shown in Figure \ref{fig:multi_links} can also be found in Figure \ref{fig:collaborations} where the authors with more survey papers are more likely to collaborate with each other.

\begin{figure}[!h]
\centering
\includegraphics[scale=0.25]{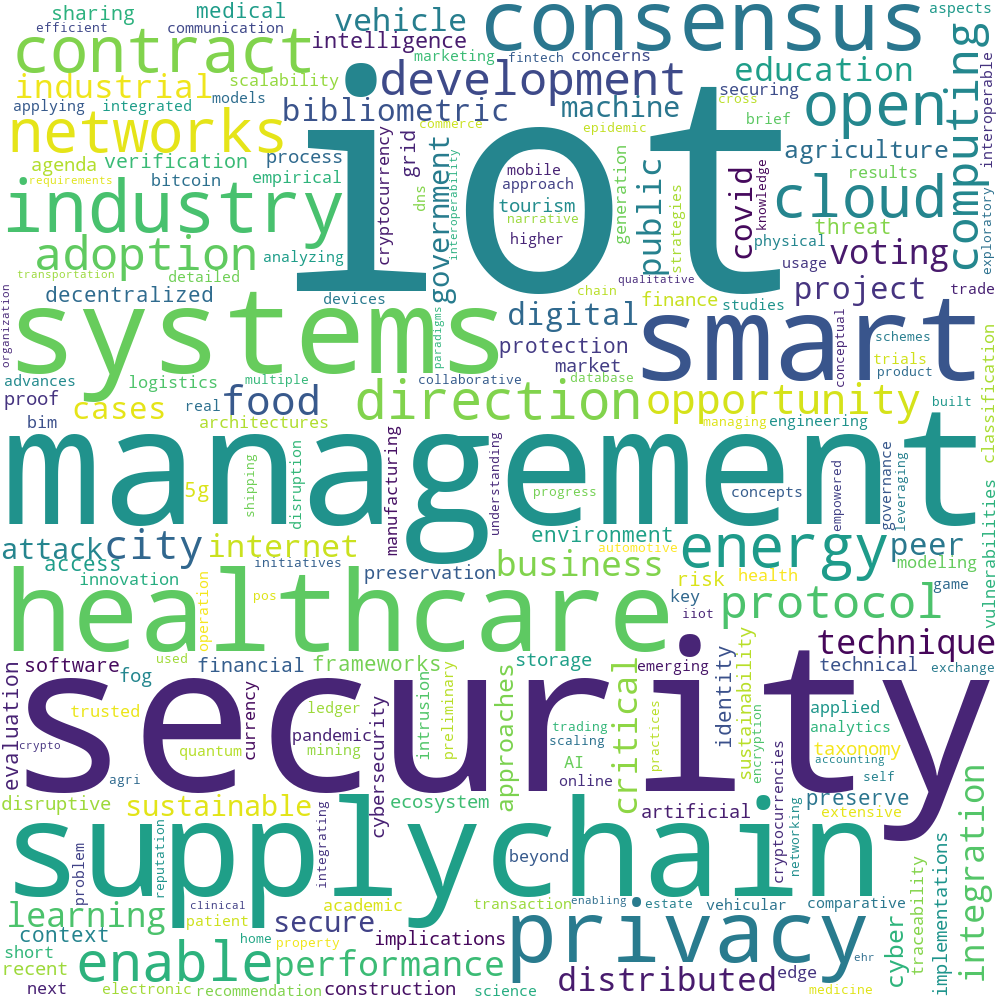}
\caption{All key words occurred in all years (January $2017$ to September $2021$).}
\label{fig:allkeywords}
\end{figure}

\subsection{Domains of highly cited papers}
We have reported the statistics on top $10$ most cited papers, their domain, and the number of citations received till September $2021$. The results are shown in Table \ref{tab:topcitations}. It can be seen that a highly cited paper is on the topic covering the ``challenges and opportunities'' of the blockchain domain. Interestingly, the second most number of citations are received by a paper which was on the topic of IoT security. Also, survey paper published in $2020$ on the topic of Blockchain security got the attention of the community and is placed on third rank with $927$ citations.

\subsection{Topic Keywords}
We extracted the keywords from the title string of all papers to identify the key domains or topics. We have removed all possible stop words like on, to, for, etc. to collect possible keywords in all papers. Also, we have combined the words which belong to the same domain, e..g, internet-of-things and all of its variants to ``IoT''. Similarly, the keywords like patient, medical, electronic medical records are replaced with ``healthcare''. The word cloud representation is shown in Figure \ref{fig:allkeywords}. A word cloud is a pictorial representation of showing the frequency of a specific word. In a word cloud, the size of a word represents word frequency, i.e., the larger the word font size, the higher the frequency. It can be from Figure \ref{fig:allkeywords} that ``IoT'' is most prominent. In fact we see from data that it occurs $164$ times. Similarly, ``security'' occurred $104$ times in the topic strings.

\subsection{Domains and Institutes}
\begin{figure}[!t]
\centering
\includegraphics[scale=0.37]{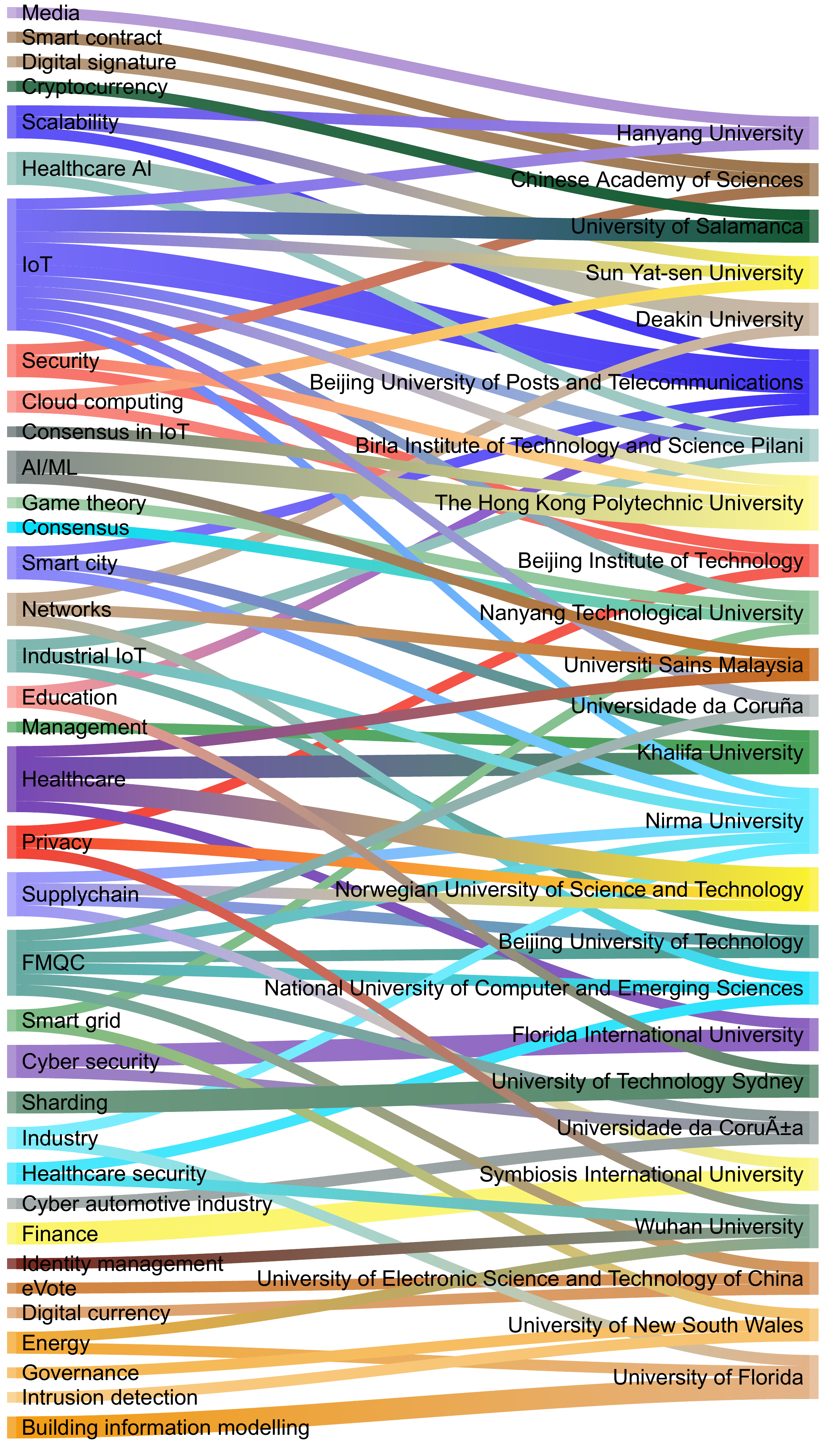}
\caption{Affiliation of domains and institutes}
\label{fig:fig12}
\end{figure}

We have identified those institutes which have worked in more than one domain and published their survey papers. To find this pattern, we concatenated the topic keywords with their respective institutes and counted the contribution of an institute against each keyword. Note that we consider the institutes of first author only. We have seen in the data that most of the institutes have worked in only one domain and published only one paper in that domain. So, just to keep the diagram simple, we selected only those institutes which have published more than two papers. The results are shown in Figure \ref{fig:fig12}. Note that the width of the link is proportional to the number of surveys published by an institute and the node width is proportional to the number of papers in a particular domain. It can be seen from Figure \ref{fig:fig12} that Beijing University of Posts and Telecommunications (BUPT) have published an overall $6$ survey papers, and its $3$ survey papers are published in the IoT domain. It can also be seen that most of the institutes have written surveys in IoT domain. We see in the data that there are $7$ institutes and each of them has published $10$ survey papers. Most of the institutes are in the Asian region and more specifically most institutes are from China and India. There is only one institute in the USA which have published $4$ survey papers in $3$ domains. Nanyang Technological University (NTU) and the University of New South Wales (UNSW) both have worked in the domain of the smart grid. University of Electronic Science and Technology of China have worked in the domain of FMQC, eVote, and digital currency, respectively. Nirma University in India has published survey papers in the most diverse domains by publishing $6$ survey papers each in IoT, supplychain, smart city, industry, industrial IoT, and FMQC.

\section{Conclusion}
\label{sec:conclusion}
We have conducted a bibliometric analysis, processed, and examined $801$ survey papers on the topic of blockchain from January $2017$ to September $2021$. We have identified the publications with respect to the publication type, publishers and venue, references, citations, paper length, different categories, year, countries, authors, and their collaborations. 

We have found that there are more journal survey papers ($510$) than conference ($242$) and not published papers ($49$). IEEE has the major percentage ($25.71$\%) of publications among the other famous publishers like Springer, Elsevier, MDPI, and ACM, which have the percentage of $13.10$\%, $12.23$\%, $6.61$\%, and $2.25$\%, respectively. Interestingly, there are more IEEE conference survey papers ($108$) than journal papers ($98$). Also, almost $55$\% of the survey papers published in Springer are conference papers. Almost $61$\% of the papers contain more than $40$ references and $72$\% of the papers have referenced the papers in the range of $10$ to $100$. Almost $30$\% of survey papers have no citations. There are $238$, $349$ and, $214$ short sized, medium-sized and large-sized papers, respectively. Statistics on countries show that $74$ countries have published the survey papers and $27$\% of the countries have only published one research paper. Most survey papers are published by India ($158$) and China ($111$), both are Asian countries and the most populated countries of the world. We also found out that there are $2556$ total authors and almost $95.18$\% of the authors have only one survey paper. Most collaborations are found to be $5$ in between author Khaled Salah and Raja Jayaraman. The most diverse and maximum number of survey papers are published with the topic keyword of ``IoT'' and ``IoT security''. Also, the second most prominent keyword was healthcare. After classifying the domains from the topic keywords, we have found that most of the institutes have worked only in one domain once. A maximum of $7$ institutes has published $10$ survey papers. There is only one institute in USA, University of Florida, has published $4$ survey papers.

\bibliographystyle{unsrtnat}
\bibliography{cas-dc-template}
\end{document}